\pdfoutput=1
\documentclass[11pt]{article}

\usepackage[]{ACL2023}

\usepackage{times}
\usepackage{latexsym}
\usepackage{subcaption}
\usepackage{comment}

\usepackage{float}

\usepackage[T1]{fontenc}
\usepackage[utf8]{inputenc}

\usepackage{microtype}

\usepackage{inconsolata}

\usepackage{lipsum} 
\usepackage{graphicx}
\usepackage{booktabs}
\usepackage{hyperref}
\usepackage{amsmath}
\usepackage{amssymb}
\usepackage{bm}
\newcommand\extrafootertext[1]{%
    \bgroup
    \renewcommand\thefootnote{\fnsymbol{footnote}}%
    \renewcommand\thempfootnote{\fnsymbol{mpfootnote}}%
    \footnotetext[0]{#1}%
    \egroup
}

\title{OLISIA: a Cascade System for Spoken Dialogue State Tracking}

\author{Léo Jacqmin$^*$\textsuperscript{1,2} \quad Lucas Druart$^*$\textsuperscript{1,3}  \\ \quad \textbf{Yannick Estève}\textsuperscript{3} 
\quad \textbf{Benoit Favre}
\textsuperscript{2} \quad \textbf{Lina M. Rojas-Barahona} \textsuperscript{1} \quad \textbf{Valentin Vielzeuf} \textsuperscript{1}\\
   \textsuperscript{1}Orange Innovation, France \\ 
    \textsuperscript{2}LIS - Aix-Marseille Université, France \\ 
    \textsuperscript{3}LIA - Avignon Université, France \\ 
    \small \textsuperscript{1}\texttt{\{leo.jacqmin,lucas1.druart,valentin.vielzeuf,linamaria.rojasbarahona\}@orange.com},\\ \small\textsuperscript{2}\texttt{\{first.last\}@lis-lab.fr},  \textsuperscript{3
}\texttt{\{first.last\}@univ-avignon.fr} 
    }

\begin{document}
\maketitle
\begin{abstract} 

Though Dialogue State Tracking (DST) is a core component of spoken dialogue systems, recent work on this task mostly deals with chat corpora, disregarding the discrepancies between spoken and written language.
In this paper, we propose OLISIA, a cascade system which integrates an Automatic Speech Recognition (ASR) model and a DST model. We introduce several adaptations in the ASR and DST modules to improve integration and robustness to spoken conversations.
With these adaptations, our system ranked first in DSTC11 Track 3, a benchmark to evaluate spoken DST. We conduct an in-depth analysis of the results and find that normalizing the ASR outputs and adapting the DST inputs through data augmentation, along with increasing the pre-trained models size all play an important role in reducing the performance discrepancy between written and spoken conversations.\footnote{Our code is made available at \url{https://github.com/Orange-OpenSource/olisia-dstc11}.} 

\end{abstract}

\section{Introduction}
\extrafootertext{$^*$ Equal contribution.}
A majority of recent research on task-oriented dialogue (TOD) systems has focused on chat corpora such as MultiWOZ \cite{budzianowskiMultiWOZLargeScaleMultiDomain2018}. With voice assistants becoming more prominent in our daily lives, there has been a renewed interest in spoken dialogue systems \cite{faruqui-tur-2022}. However, state-of-the-art systems trained on chats face robustness issues when dealing with spoken inputs \citep{kimHowRobustEvaluating2021a}. 

\begin{figure}[ht]
\centering
\includegraphics[width=1\linewidth]{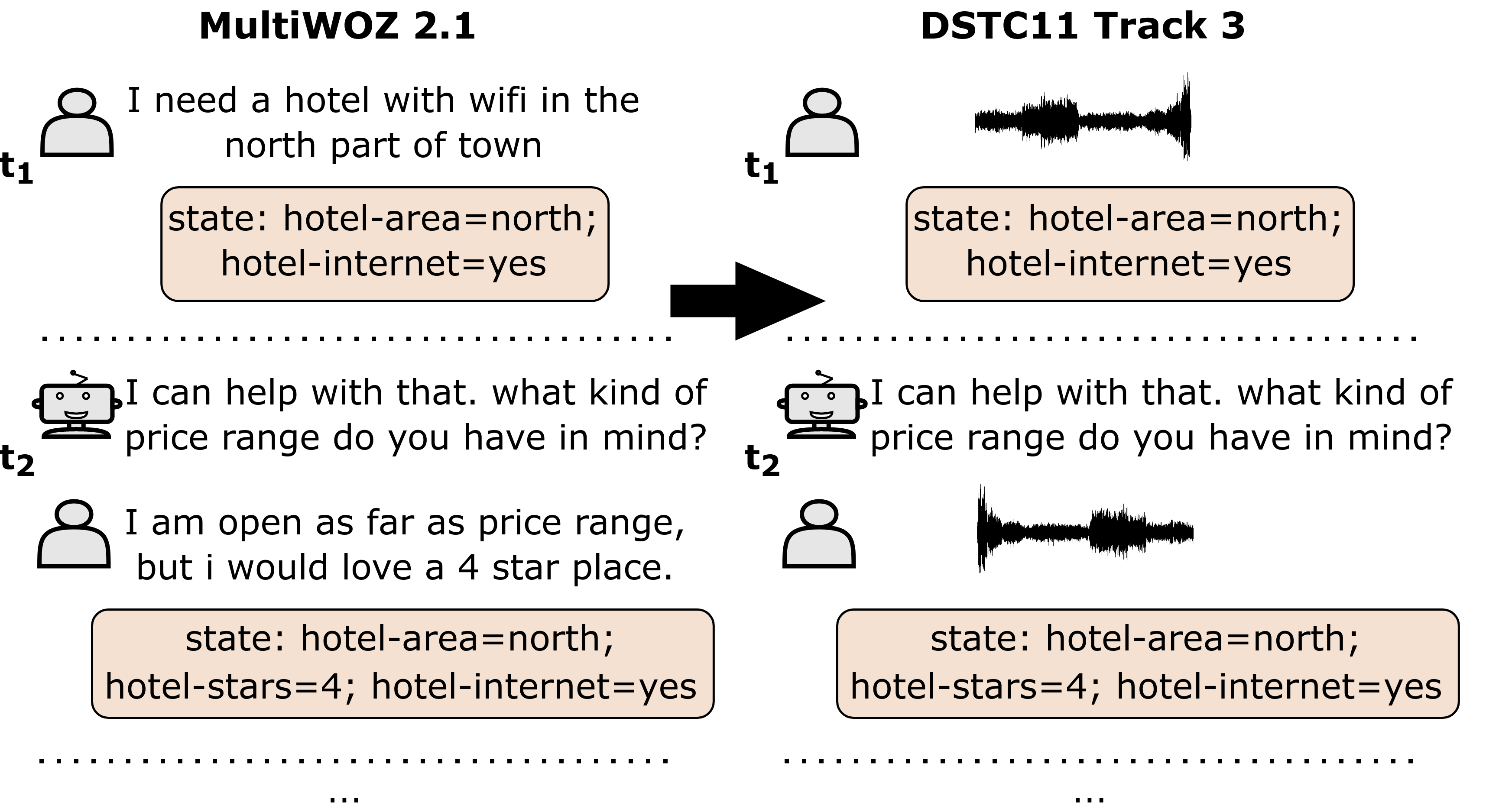}
\caption{DSTC11 Track 3 introduced a spoken version of MultiWOZ 2.1 \cite{ericMultiWOZConsolidatedMultiDomain2019} with user utterances voiced by crowdworkers.}
\label{fig:dst}
\end{figure}

In a TOD system, the role of DST is to predict at each turn and based on the dialogue history the current belief state, \textit{i.e.} a condensed and updated representation of the user needs. DST plays a central role as the system relies on the belief state to decide which action to take next. The belief state is typically frame-based and represented as a list of <slot, value> pairs. 

While both cascade and end-to-end approaches have been well studied for Spoken Language Understanding (SLU, \citet{E2ESLU}), there has been little recent work on spoken DST. Considering the entire dialogue context, as opposed to only the current turn, requires tricky strategies for end-to-end systems \cite{tomashenko2020dialogue}. In order to leverage state-of-the-art models, a cascade approach with separate ASR and DST components is thus preferred. However, these two components do not benefit from joint optimization and often lack integration.

To address these shortcomings, we propose OLISIA, a cascade system composed of an ASR model and a DST model.
OLISIA integrates these two components through adaptations in the ASR outputs and the DST inputs. 
With this design, our system ranked first in the Speech-Aware Dialog Systems Technology Challenge (DSTC11 Track~3),\footnote{\url{https://storage.googleapis.com/gresearch/dstc11/dstc11_20221102a.html}} a benchmark to evaluate spoken DST models.

In this paper, we describe our cascade spoken DST system along with the proposed adaptations.
Additionally, we conduct an analysis of these adaptations based on the results from various evaluation setups. 

Our contributions can be summarized as follows. In the context of the Speech-Aware Dialog Systems Technology Challenge, we show 
\begin{itemize}
    \item the need for post-processing the ASR output in a pipeline;
    \item the relevance of different data augmentation techniques for DST;
    \item the importance of scaling up the foundation model size -- both for ASR and DST.
\end{itemize}

\section{Related work}
\begin{figure*}[ht!]
    \centering
    \includegraphics[width=1\linewidth]{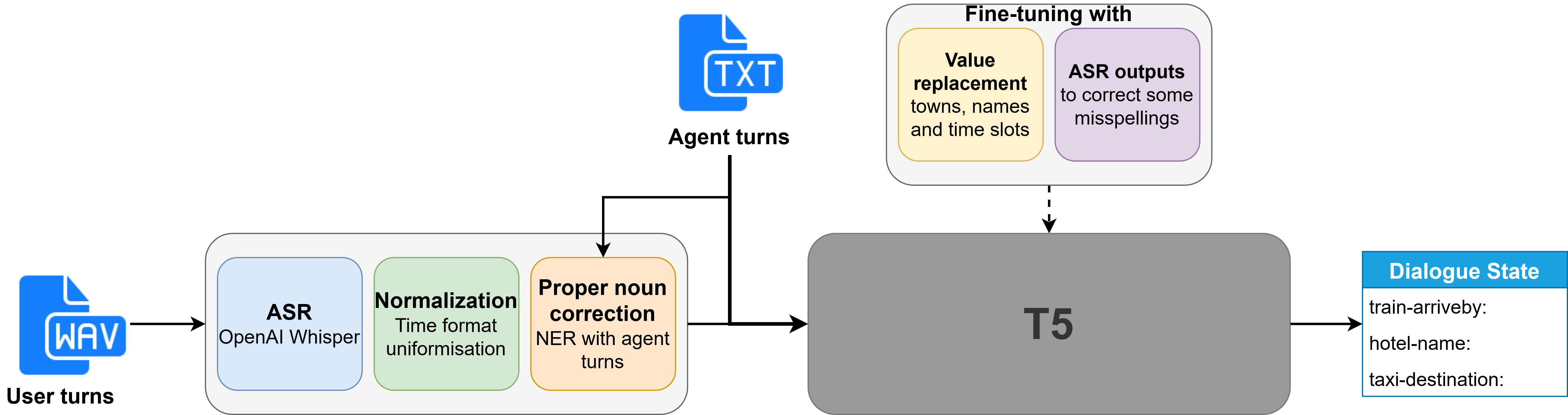}
    \caption{Illustration of OLISIA, our cascade system with adaptations of ASR and DST models to handle their respective errors.}
    \label{fig:system}
\end{figure*}

Much of the research in task-oriented dialogue (TOD) systems initially focused on spoken dialogue, for instance leveraging probabilistic modeling to account for the uncertainty associated with noisy utterances \cite{roySpokenDialogueManagement2000, williamsPartiallyObservableMarkov2007, thomsonBayesianUpdateDialogue2010}.  
The first editions of the Dialogue State Tracking Challenge\footnote{Rebranded as the Dialog Systems Technology Challenge since DSTC6.} (DSTC1 \& DSTC2, \citet{williamsDialogStateTracking2013, henderson-etal-2014-second}) introduced the first standardized benchmark for DST, releasing annotated spoken dialogue corpora. 

However, the focus of research in TOD gradually shifted to chats, assuming that the upstream ASR model would be able to provide accurate transcriptions.
Recently, there has been a renewed interest in spoken dialogue to address the lack of attention on the differences between spoken and textual inputs \cite{faruqui-tur-2022}.
DSTC10 Track~2 proposed a DST task on spoken conversations which stimulated work on this aspect, though only the $n$-best ASR hypotheses were provided without audio data.

When audio-only data is available, End-to-End Spoken Language Understanding (SLU) systems \cite{E2ESLU} are often preferred because they benefit from joint optimization. Although cascade approaches suffer from error propagation because the textual Natural Language Understanding (NLU) model does not consider the uncertainty of the ASR transcriptions, they remain competitive. In fact several adaptation techniques can boost the cascade's performance. 

Performing hypothesis rescoring with a language model specifically trained on the targeted domain proves to be effective when data is available in high quality and large quantity \cite{chung-etal-2012-lattice}. We rather adopt a post-processing approach such as spelling correction which can also help aligning the transcriptions with the targeted domain \cite{SpellingCorrection}, especially when focusing on domain-specific words. 

With the recent advances in Text-to-Speech (TTS) technologies, adapting ASR models by fine-tuning them on synthetic speech of the target domain \cite{Li2018TrainingOnTTS, ASRonTTS, ASRTTSFine-tuning} is now common. However, with only synthetic speech in the training set, such a fine-tuning might degrade the performances on human speech \cite{TTS-HumanEquilibrium}. Simulating ASR hidden representations from text in order to train an end-to-end SLU model reaches higher Named-Entity Recognition (NER) performances than training it on the synthesised speech \cite{mdhaffar22_interspeech}. Therefore, relying on the text NLU model to tolerate and correct some errors of the ASR model seems more adapted in our setting.

Recent approaches have focused on data augmentation techniques to simulate spoken data \cite{wangDataAugmentationTraining2020, liuRobustnessTestingLanguage2021, tianTODDABoostingRobustness2021} in order to make the language understanding components of TOD systems more robust to spoken inputs. Other approaches have sought to leverage the multiple hypotheses provided by the upstream ASR model in the hope that these different hypotheses complement each other to help language understanding \cite{rojas2016exploiting,liImprovingSpokenLanguage2020, ganesanNBestASRTransformer2021}. Similarly, others used a more compact representation of these hypotheses, such as word confusion networks \cite{hendersonDiscriminativeSpokenLanguage2012, palModelingASRAmbiguity2020}.

\section{Speech-Aware Dialog Systems Technology Challenge}

The lack of recent work on spoken dialogue can be attributed in part to the lack of available datasets. 
Track 3 of the Dialog Systems Technology Challenge 11\footnote{\url{https://dstc11.dstc.community/}} seeks to promote work on spoken dialogue by releasing a spoken version of MultiWOZ. This Multi-domain (restaurant, hotel, attraction, taxi, train, hospital and police) Wizard-of-Oz dataset is a large-scale human-human task-oriented conversational corpus commonly used for training and evaluating dialogue state tracking (DST), policy optimization and end-to-end dialogue modeling systems.
The goal of this track is to characterize the performance of DST models in the presence of ASR errors and speech phenomena such as disfluencies.
The organizers released a new version of MultiWOZ 2.1 \cite{ericMultiWOZConsolidatedMultiDomain2019} with user utterances voiced by crowdworkers, as illustrated in Figure \ref{fig:dst}.

Despite being widely used by the research community, MultiWOZ has been shown to exhibit an entity bias and a large overlap in the distribution of slot-values between the training and the evaluation sets which can lead to memorization in generative models \cite{qianAnnotationInconsistencyEntity2021}.
To encourage generalization, the organizers introduced modifications in the dev and test sets: the values for the slots \texttt{hotel-name}, \texttt{restaurant-name}, \texttt{train-departure} and \texttt{train-destination} were replaced with unseen entities, and time mentions were offset by a constant amount.

User utterances in the dev and test sets are vocalized by crowdworkers. A speech synthesized version of the training data is also provided in the aim of assessing the validity of such data to mitigate the lack of real spoken conversations.

Two verbatim versions of the dev set are provided to the participants, i.e. user utterances are vocalized as is by a TTS system (\textbf{TTS-v}) and human crowdworkers (\textbf{Human-v}). The test set includes the same setup along with a third version containing paraphrased user utterances vocalized by humans to sound more natural (\textbf{Human-p}).\footnote{No further details were provided regarding the value replacement and paraphrasing processes.}

System submissions are evaluated using Joint Goal Accuracy (JGA) and Slot Error Rate (SER), defined as follows:
    \[JGA =  \frac{C}{N_t} = \frac{No.\ of\ correct\ state\ pred.}{No.\ of\ turns}\]
    \[SER = \frac{S + D + I}{N_s} = \frac{No.\ of\ slot\ errors}{No.\ of\ slots\ in\ ref.}\]

where $S$, $D$ and $I$ respectively denote substitutions, deletions and insertions of <slot, value> pairs. Regarding the challenge constraints, any type of model can be used but only MultiWOZ is allowed as training data for the dialogue component.

\begin{figure*}[t!]
\centering
\includegraphics[width=\textwidth]{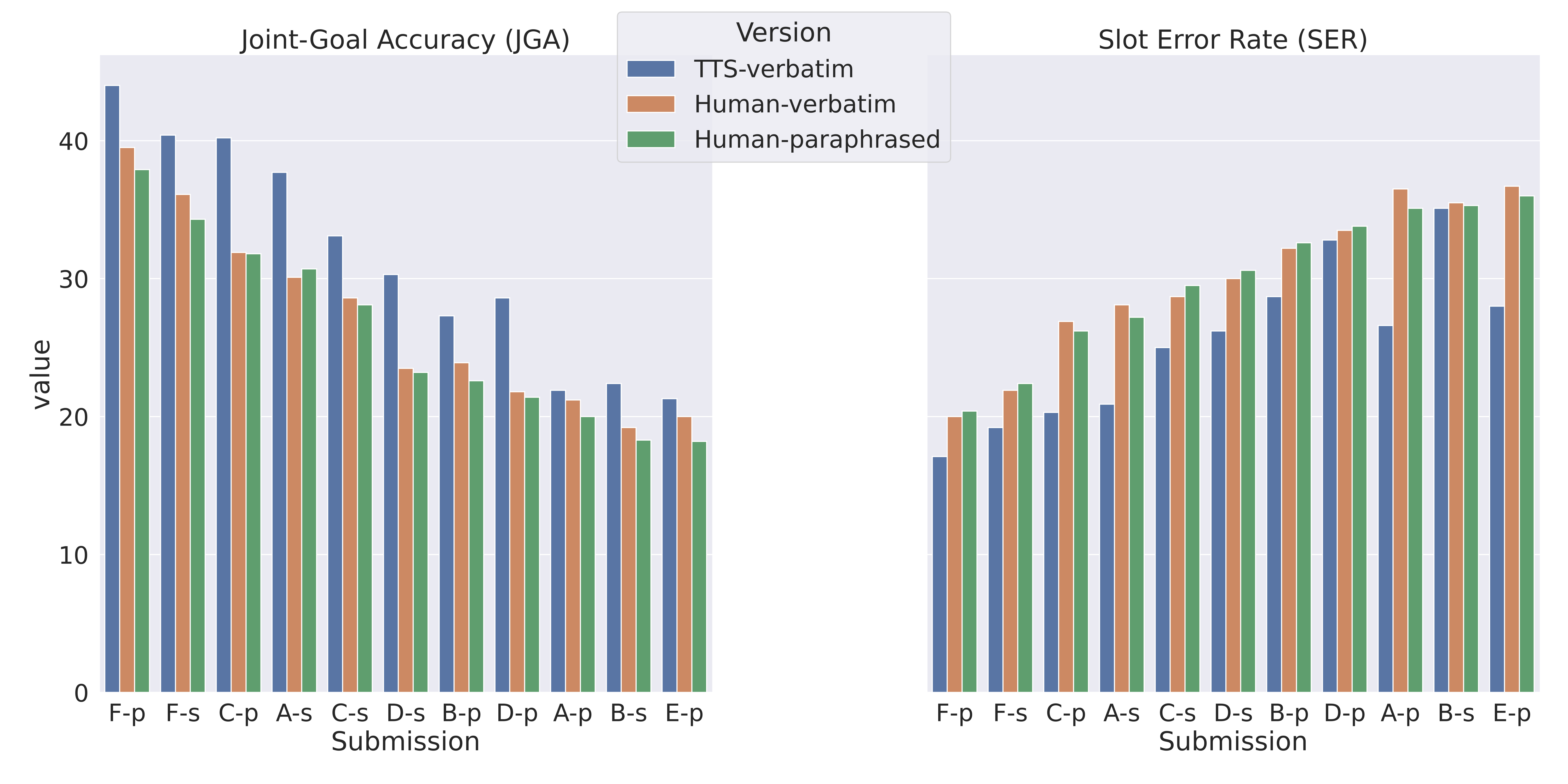}
\caption{Overall test set results (JGA$\uparrow$ / SER$\downarrow$) of the challenge for all submissions. Our primary and secondary submissions respectively correspond to \textit{F-p} and \textit{F-s}.}
\label{fig:leaderboard}
\end{figure*}

\section{Method}
\label{sec:model} 

In this section, we present our cascade approach with an ASR component which converts the user spoken inputs into text and a DST component which predicts the current dialogue state from the transcript of the previous turns. The overall architecture of the system is shown in Figure \ref{fig:system}

On the ASR side, given that the turns are perfectly segmented, we can easily transcribe the user's turns with Open AI Whisper \cite{radford22_whipser} transformer model with a forced English decoding. On the DST side, we use a generative DST model based on a pre-trained T5 model \cite{2020t5}, which proved to be more robust to spoken inputs than an extractive model in preliminary experiments. 

The input to the DST model consists of the entire dialogue history at a given turn, with agent and user utterances separated by delimiter tokens. At each turn, the model outputs the current dialogue state from scratch in the form of <slot-name, slot-value> pairs.
Formally, let $U_i$ and $A_i$ respectively be user and agent utterances at turn $i$. The input at turn $T$ is linearized by concatenating the utterances ($U_0$, $A_1$, ..., $A_{T-1}$, $U_T$) and prepending the delimiters "\texttt{user:}" and "\texttt{agent:}". The output dialogue state is linearized as a semicolon-separated list of strings "\texttt{slot-name=slot-value}".

Our contribution lies in the adaption of the transcriptions outputted from the ASR for DST and the adaptation of the DST component to handle speech specificities, which are discussed in the next two sections.

\subsection{ASR normalization}
\label{sec:ASR}

The first adaptation we apply to Whisper's transcriptions is time normalization. Given the data sources on which Whisper was trained, the outputted time formats vary a lot. We use several regular expressions to identify the most salient patterns (\textit{e.g.} "5 o' 8 am", "2 to 3 pm", "midnight", "quarter past 10 am") and map them to the standard "[hour]:[minutes] [am|pm]" format found in Multi-Woz.

The second adaptation we apply is proper noun correction which impacts the values of the slots \texttt{hotel-name}, \texttt{restaurant-name}, \texttt{taxi-destination}, \texttt{train-destination}, \texttt{taxi-departure} and \texttt{train-departure}. Many proper nouns are either misspelled by the user (\textit{e.g.} the American city Itta Bena is pronounced "I-T-T-A bena") or incorrectly recognized by Whisper (\textit{e.g.} Itta Bena transcribed as "Itta Benna"). In both cases, we use a Named Entity Recognition (NER) model\footnote{\url{https://docs.nvidia.com/deeplearning/nemo/user-guide/docs/en/v1.0.0/nlp/token_classification.html}} from Nvidia NeMo \cite{kuchaiev2019nemo} to identify lists of proper nouns from both agent and user turns. We then score each pair of user and agent identified named entities with Character Error Rate (CER) and tune a threshold in order to replace the user turns' misspelled proper nouns with their matching ones from the agent turns.

More formally, given a list of user proper nouns $l_u$, a list of agent proper nouns $l_a$ and a threshold $\delta$, we have: 
\begin{equation*}
    \forall u \in l_u, a \in l_a \ u = 
    \begin{cases}
        u \text{ if CER($u, a$) > $\delta$}\\
        a \text{ otherwise}
    \end{cases}
\end{equation*}
Where $u$ and $a$ refer respectively to user and agent proper nouns.

\begin{table*}[ht]
    \resizebox{1\textwidth}{!}{
    \begin{tabular}{lc|c|c|c|c|c|c|c|c|c}
                            & \multicolumn{4}{c}{\textbf{Dev}} & \multicolumn{6}{c}{\textbf{Test}} \\ \midrule
                            & \multicolumn{2}{c}{\textbf{TTS-v}} & \multicolumn{2}{c|}{\textbf{Human-v}} & \multicolumn{2}{c}{\textbf{TTS-v}} & \multicolumn{2}{c}{\textbf{Human-v}} & \multicolumn{2}{c}{\textbf{Human-p}} \\
         \                  & \textbf{JGA}$\uparrow$ & \textbf{SER}$\downarrow$ & \textbf{JGA}$\uparrow$ & \textbf{SER}$\downarrow$ & \textbf{JGA}$\uparrow$ & \textbf{SER}$\downarrow$ & \textbf{JGA}$\uparrow$ & \textbf{SER}$\downarrow$ & \textbf{JGA}$\uparrow$ & \textbf{SER}$\downarrow$ \\ \midrule
         \textbf{Baseline}  & 26.3 & 27.5 & 22.6 & 31.6 & \multicolumn{6}{c}{n.a.} \\
         \textbf{OLISIA$_1$}   & 47.2 & 15.7 & 43.2 & 17.9 & 44.0 & 17.1 & 39.5 & 20.0 & 37.9 & 20.4 \\ 
         \textbf{OLISIA$_2$}   & 44.1 & 17.3 & 40.3 & 19.5 & 40.4 & 19.2 & 36.0 & 21.9 & 34.3 & 22.4 \\ \midrule
         \ & \multicolumn{2}{c|}{\textbf{JGA}$\uparrow$} & \multicolumn{2}{c|}{\textbf{SER}$\downarrow$} & \multicolumn{3}{c|}{\textbf{JGA}$\uparrow$} & 
         \multicolumn{3}{c}{\textbf{SER}$\downarrow$} \\
         \textbf{OLISIA$_1$ (oracle)} & \multicolumn{2}{c|}{57.2} & \multicolumn{2}{c|}{12.5} & \multicolumn{3}{c|}{53.2} & \multicolumn{3}{c}{13.9}\\
         \textbf{OLISIA$_2$ (oracle)} & \multicolumn{2}{c|}{55.0} & \multicolumn{2}{c|}{13.6} & \multicolumn{3}{c|}{51.1} & \multicolumn{3}{c}{15.0}\\ 
        \bottomrule
    \end{tabular}
    }
    \caption{Performance (JGA$\uparrow$ / SER$\downarrow$) of our submission compared with the challenge's baseline and our system with text oracle on both dev and test sets. The baseline results on the test set were not shared.}
    \label{tab:our_results}
\end{table*}

\subsection{DST data augmentation}
\label{sec:DST}

To improve robustness and reduce the discrepancy between training and testing data, our DST model is fine-tuned on an augmented version of the provided train set. We apply value replacement, paraphrasing and speech simulation, in this order. 

In a similar way to how the dev and test sets were modified by the track organizers, our value replacement concerns named entities (town, restaurant and hotel names) and time slots. To replace entity values, we create a new ontology based on data from OpenStreetMap\footnote{\url{https://wiki.openstreetmap.org/wiki/Overpass_API}}. We then sample entities with a uniform distribution over our ontology. Values with time mentions are replaced with a random time.

The value replacement process goes as follows: 
\begin{enumerate}
    \item Successively go through each dialogue state in a dialogue, sample one value from our ontology for each distinct value and replace it in the dialogue state;
    \item Track these replacements with a mapping between replaced value and new value;
    \item Based on the obtained mapping, perform a string replacement in the dialogue context. 
\end{enumerate}
This process ensures dialogue consistency: if a value for a slot is updated during the dialogue, a new value is sampled thanks to the first step, if a value is shared between multiple slots, the same replacement value is used thanks to the second step, and finally, the string replacement in the third step is performed by decreasing lengths in order to avoid replacing sub-strings (\textit{e.g.} city names can be present in restaurant or hotel names).

Based on this new train set with replaced values, we paraphrase the user utterances using SG-GPT \cite{peng-etal-2020-shot} which allows us to condition the generation of the paraphrase on the previous turn along with the desired dialogue state for the current turn, preventing annotation inconsistencies due to hallucinations or omissions. 

Lastly, we obtain the ASR transcripts for the speech simulation by synthesizing the augmented user turns with a Tacotron2-based \cite{Shen2018NaturalTS} TTS system using SpeechBrain\footnote{\url{https://speechbrain.github.io/}} and transcribing them with Whisper.

\section{Results}
\label{sec:results}

We propose an overview of the challenge's results in section \ref{sec:overview} and further analysis of the impact of each adaptation on our system's performance in the following sections. 

\subsection{Overview}
\label{sec:overview}
The challenge's leaderboard is shown in Figure~\ref{fig:leaderboard}. All submissions have a gap between their TTS performance and human performance (14.32\%  average relative JGA decrease and 19.19\% average relative SER increase). The gap between the human-verbatim and human-paraphrased is less pronounced (3.36\%  average relative JGA decrease and 0.08\% average relative SER decrease).

Our primary submission (\textit{F-p} in Figure~\ref{fig:leaderboard}) consists in an ensemble of 5 instances of our system described in section \ref{sec:model} with transcriptions provided by Whisper-Large\footnote{\url{https://github.com/openai/whisper}} and a T5-Large model\footnote{\url{https://huggingface.co/t5-large}} fine-tuned on the variations of the training data presented in Table \ref{tab:ensemble-data}. \textbf{Replace-S} refers to one version of the train set with value replacement. With \textbf{Replace-L}, a different version of the train set is used at each epoch until convergence,\footnote{Four epochs for T5-Large.} with newly sampled entities for value replacement along with the TTS-ASR and optional paraphrasing pipeline.

We used majority vote on each predicted slot value as ensembling strategy. 
Our secondary submission (\textit{F-s} in Figure~\ref{fig:leaderboard}) consists in the best instance of the models in the ensemble (fine-tuning on $T_2$). For all models, we use a learning rate of $5e^{-4}$ and a batch size of 16. We compare their performance on the dev and test set with the challenge baseline and our system with text oracle\footnote{Text oracle considers the ground truth user turns' transcriptions. We provide these results in order to give an upper bound of our system's performances.} in Table \ref{tab:our_results}. Note that we exclude the ASR-TTS data augmentation for the system evaluated on the text oracle (+5 JGA increase).

While there is still room for improvement (10 JGA points between our system with and without text oracle), both our submissions achieved over 40 JGA on the TTS-verbatim test set. Our performance decrease (4 JGA points) from the TTS-verbatim to the Human-verbatim version is steady across the dev and test sets. We observe a smaller decrease (2 JGA points) from the Human-verbatim to the Human-paraphrased version. Finally, the difference between the dev and test sets (4 JGA points) can be explained by overfitting, a difficulty difference, or both.

\begin{table}[ht]
    \resizebox{.5\textwidth}{!}{
    \begin{tabular}{@{}lcccc@{}}
    \toprule
                     & \textbf{Replace-S} & \textbf{Replace-L} & \textbf{TTS-ASR} & \textbf{Paraphrase} \\ \midrule
    \bm{$T_1$} & \checkmark              &                &                  &                       \\
    \bm{$T_2$} & \checkmark              &                & \checkmark                &                       \\
    \bm{$T_3$} &                & \checkmark              & \checkmark                &                       \\
    \bm{$T_4$} &                & \checkmark              & \checkmark                & \checkmark                     \\
    \bm{$T_5$} &                & \checkmark$^*$              & \checkmark$^*$                &                       \\ \bottomrule
    \end{tabular}}
        \caption{Composition of the training sets used to fine-tune the models in the ensemble for our primary submission (\textbf{OLISIA$_1$}); * denotes the augmented data provided by the track organizers.}
    \label{tab:ensemble-data}
\end{table}

\subsection{ASR cascade adaptations}
\label{sec:cascade_adaptation}

Given that only a few words in the users' utterances are really important to the dialogue state, Word Error Rate (WER) is not a good measure for the quality of the transcriptions \cite{WERGoodMeasure}. Therefore we compute both JGA and WER at each correction step with the same T5-Large trained on the training set with value replacements. We present the results in Table \ref{tab:postprocessing}.  

\begin{table}[ht]
\resizebox{\linewidth}{!}{
\begin{tabular}{@{}lccc@{}}
\toprule
                                                & \textbf{TTS-v} & \textbf{Human-v} & \textbf{Human-p} \\ \midrule
\textbf{Whisper raw outputs}  & 37.9 / 4.92 & 33.8 / 8.4 & 32.0 / \_  \\
\textbf{+ Time normalization} & 40.0 / 4.49 & 35.6 / 7.89 & 33.5 / \_ \\
\textbf{+ Noun correction} & 40.3 / 4.36          & 36.1 / 7.71            & 34.3 / \_                 \\ \bottomrule
\end{tabular}}
\caption{Impact of the ASR post-processing steps on the test set performances (JGA$\uparrow$ / WER$\downarrow$).}
\label{tab:postprocessing}
\end{table}

Unsurprisingly, the TTS-verbatim version of the test set is much cleaner than the Human-verbatim version because of the higher diversity of the crowdworkers pronunciations compared to the synthetic voices. Hence we observe over 70\% relative WER increase on every correction's step outputs. This noisier version only impacts the JGA by around 10\% relative JGA decrease, confirming that only a few words matter to DST.\footnote{Note that WER values are missing for the Human-paraphrased test set as no ground truth transcripts were provided for this version.}

Time normalization improves equally all three versions of the test set (around 5\% relative JGA increase). 

Proper noun correction does not help much the clean TTS-verbatim (0.75\% relative JGA increase), however, it seems to be much more valuable for the noisier Human-verbatim and Human-paraphrased versions (respectively 1.4\% and 2.4\% relative JGA increase) again illustrating differences between synthetic and natural speech.

\subsection{Data augmentation strategies}
\label{sec:trainingset_variations}
In order to better understand how each data augmentation technique we used contributes to the overall performance, we conduct an ablation study on different versions of the training data. We incrementally add one augmentation technique at a time to the default train set and fine-tune a T5-Large model on each version. The results are shown in Table \ref{tab:dataaug}.

\begin{table}[ht]
\resizebox{\linewidth}{!}{
\begin{tabular}{@{}lccc@{}}
\toprule
                             & \textbf{TTS-v} & \textbf{Human-v} & \textbf{Human-p} \\ \midrule
\textbf{Default train set}  & 32.2                 & 28.3                    & 26.8                      \\
\textbf{+ Value replacement} & 40.2                 & 35.5                   & 33.2                      \\
\textbf{+ Speech simulation} & 40.3                  & 36.0                   & 34.3                      \\
\textbf{+ Paraphrasing}      & 37.3                 & 33.9                   & 31.5                      \\ \bottomrule
\end{tabular}}
\caption{Contribution of the different data augmentation techniques for DST in terms of JGA on the test set.}
\label{tab:dataaug}
\end{table}

\begin{figure*}[t]
\centering
\includegraphics[width=1\linewidth]{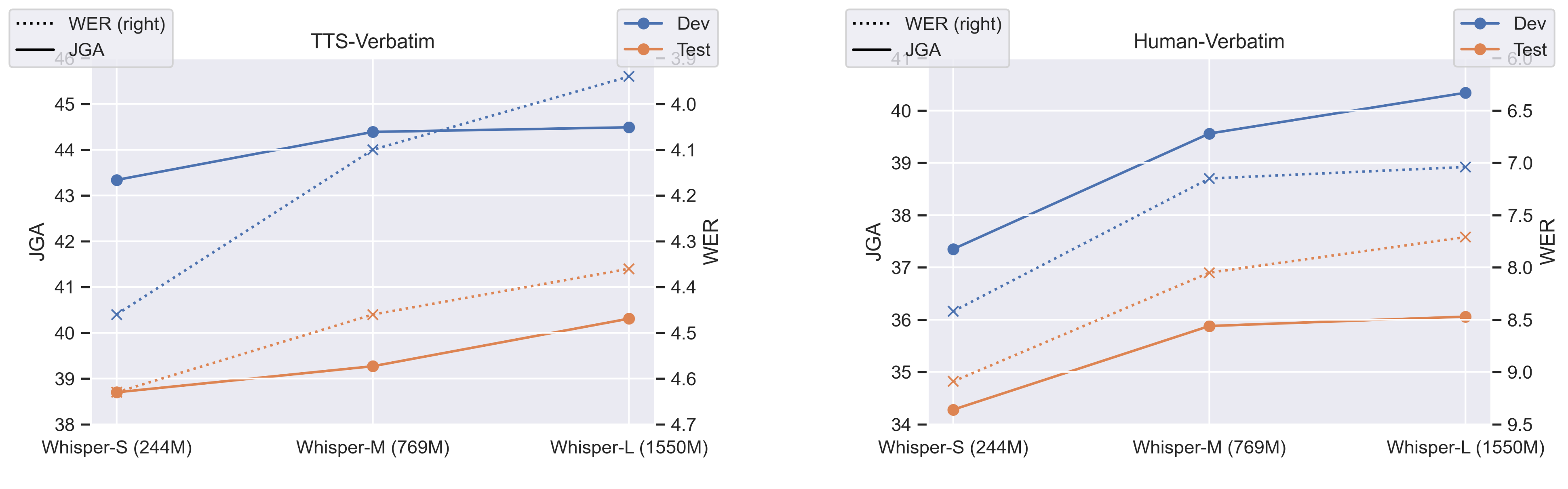}
\caption{Impact of ASR model size (no. of parameters in parenthesis) in terms of JGA$\uparrow$ in comparison with WER$\downarrow$ on both dev and test sets.}
\label{fig:whispersize}
\end{figure*}

We observe that introducing new values for the slots \texttt{hotel-name}, \texttt{restaurant-name}, \texttt{train-departure} and \texttt{train-destination} greatly alleviates the issue of memorization with MulitWOZ and enables the model to generalize more, with a 7\% to 8\% absolute increase in JGA.
The speech simulation provides an additional slight improvement which is particularly marked on the Human-paraphrased test set (+1\%). This shows that using speech synthesized training data in the absence of real spoken data can help address spoken dialogues at inference.  
On the other hand, paraphrasing the user utterances leads to an overall worse performance. One possible explanation for this result is the noisy nature of the process, using a generation model can lead to potential inconsistencies both in the flow of the dialogue and in the annotation.

\subsection{Model size}
\label{sec:model_size}

\begin{figure}[htb!]
\centering
\includegraphics[width=1\linewidth]{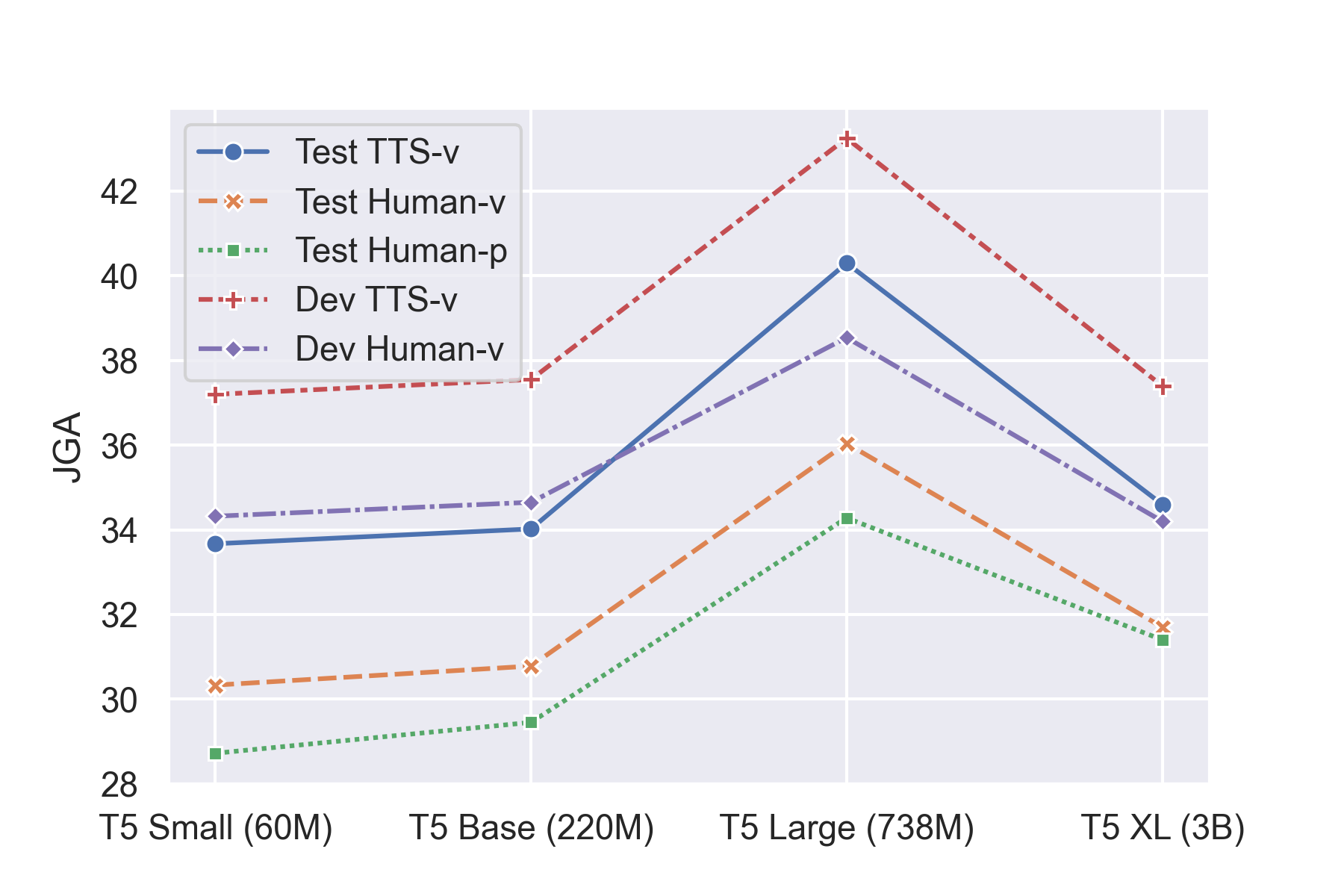}
\caption{Impact of DST model size (no. of parameters in parenthesis) in terms of JGA on the dev and test sets.}
\label{fig:dstsize}
\end{figure}

When dealing with noisy data and robustness issues we often observe that models with more parameters perform better. However, there is a trade-off between the computation resources needed for large models and the performance gains. In this section we attempt to explore this trade-off by exposing the performance gained by each model size increase. For the ASR part, we consider Whisper-Small (244M), Whisper-Medium (769M) and Whisper-Large (1550M). For the DST part, we consider T5-Small (60M), T5-Base (220M), T5-Large (738M) and T5-XL (3B), fine-tuning each model on our best training set ($T_2$).
We present the impact of the size of the ASR and DST models respectively in Figure \ref{fig:whispersize} and \ref{fig:dstsize}.

Interestingly JGA does not increase much when moving from T5-Small to T5-Base whereas it increases by almost 6 points when using T5-Large instead of T5-Base. The performance then drops back 6 points when using T5-XL, although this larger model seems to be more robust to the paraphrased test set. This suggests that XL models tend to overfit, while Large models provide a good compromise between the number of parameters and generalization. For Whisper this trend is different: using Whisper-Medium instead of Whisper-Small increases JGA of at least 2 points on the Human-verbatim dev set while using Whisper-Large instead of Whisper-Medium only increases JGA of 1 point or less. It is noteworthy that the lower parameter ratio between the Large and Medium models might explain this lower JGA increase.  

While our whisper models were not fine-tuned on any data, we can already observe that the decrease of WER obtained by using Whisper-Large instead of Whisper-Medium on the TTS-verbatim version is not found on the Human-verbatim version.

\subsection{Ensemble}
\label{sec:ensemble}
We compare two different ensembling strategies with a single model fine-tuned on $T_2$ in Table \ref{tab:ensemble_dev} (all models are based on T5-Large). Both ensembling strategies consisted of a majority vote for each slot-value. In one case, we used five models fine-tuned on the same train set with different random seeds and in the other, we fine-tuned five models on five different variations of the train set (\textit{c.f.} Table \ref{tab:ensemble-data}). 

\begin{table}[ht]
\resizebox{\linewidth}{!}{
\begin{tabular}{@{}lcc@{}}
\toprule
                                         & \textbf{TTS-v} & \textbf{Human-v} \\ \midrule
\textbf{1 model (no ensemble)}           & 44.1                 & 40.2                   \\
\textbf{5 models - same train set}       & 44.4                 & 40.4                   \\
\textbf{5 models - different train sets} & 47.8                 & 43.5                   \\ 
\textbf{9 models - different train sets} & 48.5                 & 43.9                 \\
\bottomrule
\end{tabular}}
\caption{Comparison of different ensembling strategies on the dev set in terms of JGA.}
\label{tab:ensemble_dev}
\end{table}

We find that the value of the ensemble from the same train set is limited, only providing a slight increase in JGA compared to the single model. The advantage of ensembling appears with five models fine-tuned on different train sets, providing a 3\% absolute increase in JGA. This suggests that the ensemble benefits from having contrasting views of the same instance at evaluation, and by extension that T5 models fail to learn invariant representations of proper nouns (or that our method does not enable that). It is also noteworthy that performance does not increase that much beyond 5 models, with marginally better results at 9 models,\footnote{Post-evaluation experiments} likely showing a performance ceiling. 

\section{Limitations}

Overall, the dataset released in this challenge is a good step towards bridging the gap between written and spoken dialogue systems. However, as the user utterances were read aloud by humans, this spoken data lacks in spontaneity associated with actual speech. It would be interesting to see if our findings hold on spontaneous spoken dialogue. 

One limitation of our system lies in the use of Transformer-based models for both ASR and DST. While these models provide attractive performances they come with their own limitation of quadratic memory and limited input size. Dialogues in MultiWOZ are relatively short and this was not a concern in this work, but this could become problematic when dealing with longer, more realistic conversations. One alternative would be to reduce the input context to the most recent turns and include the linearized previous dialogue state in the DST input. 

Another concern is the use of large pre-trained models to achieve competitive results. As pointed out before \cite{strubell-etal-2019-energy}, training these large models on abundant data requires a substantial electrical consumption. In light of the human impact on the environment, we should promote efficiency as the main performance factor rather than metric scores. 

\section{Conclusion and future work}
This work introduced OLISIA, a cascade system for spoken DST that integrates an ASR and a DST model through several adaptations. We used ASR normalization and DST data augmentation to adjust each component to its counterpart. We have shown the importance of these adaptations in improving robustness to spoken inputs and our system achieved first place in the Speech-Aware Dialog Systems Technology Challenge.

While having user turns as speech and agent turns as text is a natural setup for a spoken dialogue system, this mix of modalities makes it more challenging to develop end-to-end systems. Recent progress on speech and text multimodal models could prove to be useful in addressing this problem \cite{ao-etal-2022-speecht5}. Another possibility would be to exploit agent turns using an audio only model, either as synthesized speech or as intermediary ASR representations \cite{mdhaffar22_interspeech}.

This work focused on a cascade approach but a thorough comparison of end-to-end and cascade approaches could also be helpful in further research on spoken dialogue systems.

\section*{Acknowledgements}
We thank Marco Dinareli for the valuable discussions and fruitful suggestions regarding end-to-end spoken DST.

This work was partially funded by the ANR CIFRE contracts N°2022/0286 and N°2022/0011 and by Orange Innovation.

This work was performed using HPC resources from GENCI-IDRIS (Grant 2022-AD011011407R2).

\bibliographystyle{acl_natbib}
\bibliography{custom}

\begin{thebibliography}{35}
\expandafter\ifx\csname natexlab\endcsname\relax\def\natexlab#1{#1}\fi

\bibitem[{Ao et~al.(2022)Ao, Wang, Zhou, Wang, Ren, Wu, Liu, Ko, Li, Zhang,
  Wei, Qian, Li, and Wei}]{ao-etal-2022-speecht5}
Junyi Ao, Rui Wang, Long Zhou, Chengyi Wang, Shuo Ren, Yu~Wu, Shujie Liu, Tom
  Ko, Qing Li, Yu~Zhang, Zhihua Wei, Yao Qian, Jinyu Li, and Furu Wei. 2022.
\newblock \href {https://doi.org/10.18653/v1/2022.acl-long.393} {{S}peech{T}5:
  Unified-modal encoder-decoder pre-training for spoken language processing}.
\newblock In \emph{Proceedings of the 60th Annual Meeting of the Association
  for Computational Linguistics (Volume 1: Long Papers)}, pages 5723--5738,
  Dublin, Ireland. Association for Computational Linguistics.

\bibitem[{Budzianowski et~al.(2018)Budzianowski, Wen, Tseng, Casanueva, Ultes,
  Ramadan, and Ga{\v s}i{\'c}}]{budzianowskiMultiWOZLargeScaleMultiDomain2018}
Pawe{\textbackslash}l Budzianowski, Tsung-Hsien Wen, Bo-Hsiang Tseng, I{\~n}igo
  Casanueva, Stefan Ultes, Osman Ramadan, and Milica Ga{\v s}i{\'c}. 2018.
\newblock {{MultiWOZ}} - {{A Large-Scale Multi-Domain Wizard-of-Oz Dataset}}
  for {{Task-Oriented Dialogue Modelling}}.
\newblock In \emph{Proceedings of the 2018 {{Conference}} on {{Empirical
  Methods}} in {{Natural Language Processing}}}, pages 5016--5026, {Brussels,
  Belgium}. {Association for Computational Linguistics}.

\bibitem[{Chung et~al.(2012)Chung, Jeon, Park, and
  Lee}]{chung-etal-2012-lattice}
Euisok Chung, Hyung-Bae Jeon, Jeon-Gue Park, and Yun-Keun Lee. 2012.
\newblock \href {https://aclanthology.org/C12-2022} {Lattice rescoring for
  speech recognition using large scale distributed language models}.
\newblock In \emph{Proceedings of {COLING} 2012: Posters}, pages 217--224,
  Mumbai, India. The COLING 2012 Organizing Committee.

\bibitem[{Eric et~al.(2019)Eric, Goel, Paul, Kumar, Sethi, Ku, Goyal, Agarwal,
  Gao, and {Hakkani-Tur}}]{ericMultiWOZConsolidatedMultiDomain2019}
Mihail Eric, Rahul Goel, Shachi Paul, Adarsh Kumar, Abhishek Sethi, Peter Ku,
  Anuj~Kumar Goyal, Sanchit Agarwal, Shuyang Gao, and Dilek {Hakkani-Tur}.
  2019.
\newblock \href {http://arxiv.org/abs/1907.01669} {{{MultiWOZ}} 2.1: {{A
  Consolidated Multi-Domain Dialogue Dataset}} with {{State Corrections}} and
  {{State Tracking Baselines}}}.
\newblock \emph{arXiv:1907.01669 [cs]}.

\bibitem[{Faruqui and Hakkani-Tür(2022)}]{faruqui-tur-2022}
Manaal Faruqui and Dilek Hakkani-Tür. 2022.
\newblock \href
  {http://arxiv.org/abs/https://direct.mit.edu/coli/article-pdf/48/1/221/2006612/coli\_a\_00430.pdf}
  {{Revisiting the Boundary between ASR and NLU in the Age of Conversational
  Dialog Systems}}.
\newblock \emph{Computational Linguistics}, 48(1):221--232.

\bibitem[{Ganesan et~al.(2021)Ganesan, Bamdev, B, Venugopal, and
  Tushar}]{ganesanNBestASRTransformer2021}
Karthik Ganesan, Pakhi Bamdev, Jaivarsan B, Amresh Venugopal, and Abhinav
  Tushar. 2021.
\newblock \href {https://doi.org/10.18653/v1/2021.acl-short.14} {N-{{Best ASR
  Transformer}}: {{Enhancing SLU Performance}} using {{Multiple ASR
  Hypotheses}}}.
\newblock In \emph{Proceedings of the 59th {{Annual Meeting}} of the
  {{Association}} for {{Computational Linguistics}} and the 11th
  {{International Joint Conference}} on {{Natural Language Processing}}
  ({{Volume}} 2: {{Short Papers}})}, pages 93--98, {Online}. {Association for
  Computational Linguistics}.

\bibitem[{Henderson et~al.(2012)Henderson, Ga{\v s}i{\'c}, Thomson, Tsiakoulis,
  Yu, and Young}]{hendersonDiscriminativeSpokenLanguage2012}
Matthew Henderson, Milica Ga{\v s}i{\'c}, Blaise Thomson, Pirros Tsiakoulis,
  Kai Yu, and Steve Young. 2012.
\newblock \href {https://doi.org/10.1109/SLT.2012.6424218} {Discriminative
  spoken language understanding using word confusion networks}.
\newblock In \emph{2012 {{IEEE Spoken Language Technology Workshop}}
  ({{SLT}})}, pages 176--181.

\bibitem[{Henderson et~al.(2014)Henderson, Thomson, and
  Williams}]{henderson-etal-2014-second}
Matthew Henderson, Blaise Thomson, and Jason~D. Williams. 2014.
\newblock \href {https://doi.org/10.3115/v1/W14-4337} {The second dialog state
  tracking challenge}.
\newblock In \emph{Proceedings of the 15th Annual Meeting of the Special
  Interest Group on Discourse and Dialogue ({SIGDIAL})}, pages 263--272,
  Philadelphia, PA, U.S.A. Association for Computational Linguistics.

\bibitem[{Hrinchuk et~al.(2020)Hrinchuk, Popova, and
  Ginsburg}]{SpellingCorrection}
Oleksii Hrinchuk, Mariya Popova, and Boris Ginsburg. 2020.
\newblock \href {https://doi.org/10.1109/ICASSP40776.2020.9053051} {Correction
  of automatic speech recognition with transformer sequence-to-sequence model}.
\newblock In \emph{ICASSP 2020 - 2020 IEEE International Conference on
  Acoustics, Speech and Signal Processing (ICASSP)}, pages 7074--7078.

\bibitem[{Kim et~al.(2021)Kim, Liu, Jin, Papangelis, Gopalakrishnan,
  Hedayatnia, and {Hakkani-T{\"u}r}}]{kimHowRobustEvaluating2021a}
Seokhwan Kim, Yang Liu, Di~Jin, A.~Papangelis, Karthik Gopalakrishnan, Behnam
  Hedayatnia, and Dilek~Z. {Hakkani-T{\"u}r}. 2021.
\newblock ``{{How Robust R U}}?'': {{Evaluating Task-Oriented Dialogue
  Systems}} on {{Spoken Conversations}}.
\newblock \emph{2021 IEEE Automatic Speech Recognition and Understanding
  Workshop (ASRU)}.

\bibitem[{Kuchaiev et~al.(2019)Kuchaiev, Li, Nguyen, Hrinchuk, Leary, Ginsburg,
  Kriman, Beliaev, Lavrukhin, Cook et~al.}]{kuchaiev2019nemo}
Oleksii Kuchaiev, Jason Li, Huyen Nguyen, Oleksii Hrinchuk, Ryan Leary, Boris
  Ginsburg, Samuel Kriman, Stanislav Beliaev, Vitaly Lavrukhin, Jack Cook,
  et~al. 2019.
\newblock Nemo: a toolkit for building ai applications using neural modules.
\newblock \emph{arXiv preprint arXiv:1909.09577}.

\bibitem[{Laptev et~al.(2020)Laptev, Korostik, Svischev, Andrusenko,
  Medennikov, and Rybin}]{TTS-HumanEquilibrium}
Aleksandr Laptev, Roman Korostik, Aleksey Svischev, Andrei Andrusenko, Ivan
  Medennikov, and Sergey Rybin. 2020.
\newblock \href {https://doi.org/10.1109/CISP-BMEI51763.2020.9263564} {You do
  not need more data: Improving end-to-end speech recognition by text-to-speech
  data augmentation}.
\newblock In \emph{2020 13th International Congress on Image and Signal
  Processing, BioMedical Engineering and Informatics (CISP-BMEI)}, pages
  439--444.

\bibitem[{Li et~al.(2018)Li, Gadde, Ginsburg, and
  Lavrukhin}]{Li2018TrainingOnTTS}
Jason Li, Ravi~Teja Gadde, Boris Ginsburg, and Vitaly Lavrukhin. 2018.
\newblock Training neural speech recognition systems with synthetic speech
  augmentation.
\newblock \emph{ArXiv}, abs/1811.00707.

\bibitem[{Li et~al.(2020)Li, Ruan, Liu, Soldaini, Hamza, and
  Su}]{liImprovingSpokenLanguage2020}
Mingda Li, Weitong Ruan, Xinyue Liu, Luca Soldaini, Wael Hamza, and Chengwei
  Su. 2020.
\newblock \href {http://arxiv.org/abs/2001.05284} {Improving {{Spoken Language
  Understanding By Exploiting ASR N-best Hypotheses}}}.
\newblock \emph{arXiv:2001.05284 [cs, eess]}.

\bibitem[{Liu et~al.(2021)Liu, Takanobu, Wen, Wan, Li, Nie, Li, Peng, and
  Huang}]{liuRobustnessTestingLanguage2021}
Jiexi Liu, Ryuichi Takanobu, Jiaxin Wen, Dazhen Wan, Hongguang Li, Weiran Nie,
  Cheng Li, Wei Peng, and Minlie Huang. 2021.
\newblock \href {https://doi.org/10.18653/v1/2021.acl-long.192} {Robustness
  {{Testing}} of {{Language Understanding}} in {{Task-Oriented Dialog}}}.
\newblock In \emph{Proceedings of the 59th {{Annual Meeting}} of the
  {{Association}} for {{Computational Linguistics}} and the 11th
  {{International Joint Conference}} on {{Natural Language Processing}}
  ({{Volume}} 1: {{Long Papers}})}, pages 2467--2480, {Online}. {Association
  for Computational Linguistics}.

\bibitem[{Mdhaffar et~al.(2022)Mdhaffar, Duret, Parcollet, and
  Estève}]{mdhaffar22_interspeech}
Salima Mdhaffar, Jarod Duret, Titouan Parcollet, and Yannick Estève. 2022.
\newblock \href {https://doi.org/10.21437/Interspeech.2022-10231} {{End-to-end
  model for named entity recognition from speech without paired training
  data}}.
\newblock In \emph{Proc. Interspeech 2022}, pages 4068--4072.

\bibitem[{Pal et~al.(2020)Pal, Guillot, Shrivastava, Renders, and
  Besacier}]{palModelingASRAmbiguity2020}
Vaishali Pal, Fabien Guillot, Manish Shrivastava, Jean-Michel Renders, and
  Laurent Besacier. 2020.
\newblock \href {https://hal.archives-ouvertes.fr/hal-02962177} {Modeling {{ASR
  Ambiguity}} for {{Neural Dialogue State Tracking}}}.
\newblock In \emph{Interspeech 2020}, {Shangai (Virtual Conf), China}.

\bibitem[{Peng et~al.(2020)Peng, Zhu, Li, Li, Li, Zeng, and
  Gao}]{peng-etal-2020-shot}
Baolin Peng, Chenguang Zhu, Chunyuan Li, Xiujun Li, Jinchao Li, Michael Zeng,
  and Jianfeng Gao. 2020.
\newblock \href {https://doi.org/10.18653/v1/2020.findings-emnlp.17} {Few-shot
  natural language generation for task-oriented dialog}.
\newblock In \emph{Findings of the Association for Computational Linguistics:
  EMNLP 2020}, pages 172--182, Online. Association for Computational
  Linguistics.

\bibitem[{Qian et~al.(2021)Qian, Beirami, Lin, De, Geramifard, Yu, and
  Sankar}]{qianAnnotationInconsistencyEntity2021}
Kun Qian, Ahmad Beirami, Zhouhan Lin, Ankita De, Alborz Geramifard, Zhou Yu,
  and Chinnadhurai Sankar. 2021.
\newblock \href {https://aclanthology.org/2021.sigdial-1.35} {Annotation
  {{Inconsistency}} and {{Entity Bias}} in {{MultiWOZ}}}.
\newblock In \emph{Proceedings of the 22nd {{Annual Meeting}} of the {{Special
  Interest Group}} on {{Discourse}} and {{Dialogue}}}, pages 326--337,
  {Singapore and Online}. {Association for Computational Linguistics}.

\bibitem[{Radford et~al.(2022)Radford, Kim, Xu, Brockman, McLeavey, and
  Sutskever}]{radford22_whipser}
Alec Radford, Jong~Wook Kim, Tao Xu, Greg Brockman, Christine McLeavey, and
  Ilya Sutskever. 2022.
\newblock \href {https://cdn.openai.com/papers/whisper.pdf} {{Robust Speech
  Recognition via Large-Scale Weak Supervision}}.
\newblock Technical report, OpenAI.

\bibitem[{Raffel et~al.(2020)Raffel, Shazeer, Roberts, Lee, Narang, Matena,
  Zhou, Li, and Liu}]{2020t5}
Colin Raffel, Noam Shazeer, Adam Roberts, Katherine Lee, Sharan Narang, Michael
  Matena, Yanqi Zhou, Wei Li, and Peter~J. Liu. 2020.
\newblock \href {http://jmlr.org/papers/v21/20-074.html} {Exploring the limits
  of transfer learning with a unified text-to-text transformer}.
\newblock \emph{Journal of Machine Learning Research}, 21(140):1--67.

\bibitem[{Rojas-Barahona et~al.(2016)Rojas-Barahona, Gasic, Mrk{\v{s}}i{\'c},
  Su, Ultes, Wen, and Young}]{rojas2016exploiting}
Lina~M. Rojas-Barahona, Milica Gasic, Nikola Mrk{\v{s}}i{\'c}, Pei-Hao Su,
  Stefan Ultes, Tsung-Hsien Wen, and Steve Young. 2016.
\newblock Exploiting sentence and context representations in deep neural models
  for spoken language understanding.
\newblock \emph{arXiv preprint arXiv:1610.04120}.

\bibitem[{Rosenberg et~al.(2019)Rosenberg, Zhang, Ramabhadran, Jia, Moreno, Wu,
  and Wu}]{ASRonTTS}
Andrew Rosenberg, Yu~Zhang, Bhuvana Ramabhadran, Ye~Jia, Pedro Moreno, Yonghui
  Wu, and Zelin Wu. 2019.
\newblock \href {https://doi.org/10.1109/ASRU46091.2019.9003990} {Speech
  recognition with augmented synthesized speech}.
\newblock In \emph{2019 IEEE Automatic Speech Recognition and Understanding
  Workshop (ASRU)}, pages 996--1002.

\bibitem[{Roy et~al.(2000)Roy, Pineau, and
  Thrun}]{roySpokenDialogueManagement2000}
Nicholas Roy, Joelle Pineau, and Sebastian Thrun. 2000.
\newblock \href {https://doi.org/10.3115/1075218.1075231} {Spoken dialogue
  management using probabilistic reasoning}.
\newblock In \emph{Proceedings of the 38th {{Annual Meeting}} on
  {{Association}} for {{Computational Linguistics}} - {{ACL}} '00}, pages
  93--100, {Hong Kong}. {Association for Computational Linguistics}.

\bibitem[{Serdyuk et~al.(2018)Serdyuk, Wang, Fuegen, Kumar, Liu, and
  Bengio}]{E2ESLU}
Dmitriy Serdyuk, Yongqiang Wang, Christian Fuegen, Anuj Kumar, Baiyang Liu, and
  Yoshua Bengio. 2018.
\newblock \href {https://doi.org/10.1109/ICASSP.2018.8461785} {Towards
  end-to-end spoken language understanding}.
\newblock In \emph{2018 IEEE International Conference on Acoustics, Speech and
  Signal Processing (ICASSP)}, pages 5754--5758.

\bibitem[{Shen et~al.(2018)Shen, Pang, Weiss, Schuster, Jaitly, Yang, Chen,
  Zhang, Wang, Skerry-Ryan, Saurous, Agiomyrgiannakis, and
  Wu}]{Shen2018NaturalTS}
Jonathan Shen, Ruoming Pang, Ron~J. Weiss, Mike Schuster, Navdeep Jaitly,
  Zongheng Yang, Z.~Chen, Yu~Zhang, Yuxuan Wang, R.~J. Skerry-Ryan, Rif~A.
  Saurous, Yannis Agiomyrgiannakis, and Yonghui Wu. 2018.
\newblock Natural tts synthesis by conditioning wavenet on mel spectrogram
  predictions.
\newblock \emph{2018 IEEE International Conference on Acoustics, Speech and
  Signal Processing (ICASSP)}, pages 4779--4783.

\bibitem[{Strubell et~al.(2019)Strubell, Ganesh, and
  McCallum}]{strubell-etal-2019-energy}
Emma Strubell, Ananya Ganesh, and Andrew McCallum. 2019.
\newblock \href {https://doi.org/10.18653/v1/P19-1355} {Energy and policy
  considerations for deep learning in {NLP}}.
\newblock In \emph{Proceedings of the 57th Annual Meeting of the Association
  for Computational Linguistics}, pages 3645--3650, Florence, Italy.
  Association for Computational Linguistics.

\bibitem[{Thomson and Young(2010)}]{thomsonBayesianUpdateDialogue2010}
Blaise Thomson and Steve Young. 2010.
\newblock \href {https://doi.org/10.1016/j.csl.2009.07.003} {Bayesian update of
  dialogue state: {{A POMDP}} framework for spoken dialogue systems}.
\newblock \emph{Computer Speech \& Language}, 24(4):562--588.

\bibitem[{Tian et~al.(2021)Tian, Huang, He, Lin, Bao, He, Huang, Ju, Zhang,
  Xie, Sun, Wang, Wu, and Wang}]{tianTODDABoostingRobustness2021}
Xin Tian, Xinxian Huang, Dongfeng He, Yingzhan Lin, Siqi Bao, Huang He, Liankai
  Huang, Qiang Ju, Xiyuan Zhang, Jian Xie, Shuqi Sun, Fan Wang, Hua Wu, and
  Haifeng Wang. 2021.
\newblock \href {http://arxiv.org/abs/2112.12441} {{{TOD-DA}}: {{Towards
  Boosting}} the {{Robustness}} of {{Task-oriented Dialogue Modeling}} on
  {{Spoken Conversations}}}.
\newblock \emph{arXiv:2112.12441 [cs]}.

\bibitem[{Tomashenko et~al.(2020)Tomashenko, Raymond, Caubri{\`e}re, De~Mori,
  and Est{\`e}ve}]{tomashenko2020dialogue}
Natalia Tomashenko, Christian Raymond, Antoine Caubri{\`e}re, Renato De~Mori,
  and Yannick Est{\`e}ve. 2020.
\newblock Dialogue history integration into end-to-end signal-to-concept spoken
  language understanding systems.
\newblock In \emph{ICASSP 2020-2020 IEEE International Conference on Acoustics,
  Speech and Signal Processing (ICASSP)}, pages 8509--8513. IEEE.

\bibitem[{Wang et~al.(2020)Wang, {Fazel-Zarandi}, Tiwari, Matsoukas, and
  Polymenakos}]{wangDataAugmentationTraining2020}
Longshaokan Wang, Maryam {Fazel-Zarandi}, Aditya Tiwari, Spyros Matsoukas, and
  Lazaros Polymenakos. 2020.
\newblock \href {https://doi.org/10.18653/v1/2020.nlp4convai-1.8} {Data
  {{Augmentation}} for {{Training Dialog Models Robust}} to {{Speech
  Recognition Errors}}}.
\newblock In \emph{Proceedings of the 2nd {{Workshop}} on {{Natural Language
  Processing}} for {{Conversational AI}}}, pages 63--70, {Online}. {Association
  for Computational Linguistics}.

\bibitem[{Wang et~al.(2003)Wang, Acero, and Chelba}]{WERGoodMeasure}
Ye-Yi Wang, A.~Acero, and C.~Chelba. 2003.
\newblock \href {https://doi.org/10.1109/ASRU.2003.1318504} {Is word error rate
  a good indicator for spoken language understanding accuracy}.
\newblock In \emph{2003 IEEE Workshop on Automatic Speech Recognition and
  Understanding (IEEE Cat. No.03EX721)}, pages 577--582.

\bibitem[{Williams et~al.(2013)Williams, Raux, Ramachandran, and
  Black}]{williamsDialogStateTracking2013}
Jason Williams, Antoine Raux, Deepak Ramachandran, and Alan Black. 2013.
\newblock \href {https://aclanthology.org/W13-4065} {The {{Dialog State
  Tracking Challenge}}}.
\newblock In \emph{Proceedings of the {{SIGDIAL}} 2013 {{Conference}}}, pages
  404--413, {Metz, France}. {Association for Computational Linguistics}.

\bibitem[{Williams and Young(2007)}]{williamsPartiallyObservableMarkov2007}
Jason~D. Williams and Steve Young. 2007.
\newblock \href {https://doi.org/10.1016/j.csl.2006.06.008} {Partially
  observable {{Markov}} decision processes for spoken dialog systems}.
\newblock \emph{Computer Speech \& Language}, 21(2):393--422.

\bibitem[{Zheng et~al.(2021)Zheng, Liu, Gunceler, and
  Willett}]{ASRTTSFine-tuning}
Xianrui Zheng, Yulan Liu, Deniz Gunceler, and Daniel Willett. 2021.
\newblock \href {https://doi.org/10.1109/ICASSP39728.2021.9414778} {Using
  synthetic audio to improve the recognition of out-of-vocabulary words in
  end-to-end asr systems}.
\newblock In \emph{ICASSP 2021 - 2021 IEEE International Conference on
  Acoustics, Speech and Signal Processing (ICASSP)}, pages 5674--5678.

\end{thebibliography}

\end{document}